\documentclass[a4paper,prc,twocolumn]{revtex4-1}
\usepackage{amsmath,graphicx}
\begin{document}
\title{Elliptic flow at energies available at the CERN Large Hadron Collider:  Comparing heavy-ion data to viscous hydrodynamic predictions}
\author{Matthew Luzum}
\affiliation{
CEA, Institut de physique th\'eorique de Saclay (IPhT), F-91191
Gif-sur-Yvette, France}
%\author{Paul Romatschke}
%\affiliation{Frankfurt Institute for Advanced Studies, D-60438 Frankfurt, Germany}
%\date{\today}
\begin{abstract}
I compare the first viscous hydrodynamic prediction for integrated elliptic flow in Pb-Pb collisions at the LHC with the first data released by the ALICE collaboration.  These new data are found to be consistent with hydrodynamic extrapolations of RHIC data with no change in medium parameters (e.g., average viscosity).  I also discuss how, in general, a precise comparison of data to theoretical calculations requires an understanding of some subtleties of the measurement --- most notably the cut on transverse momentum of the particles used and the differing sensitivities to flow fluctuations and non-flow effects of the various measurement methods.
\end{abstract}
\maketitle
%
%
%
%\section{Introduction}
The ALICE collaboration recently released the first results from heavy ion collisions at the Large Hadron Collider (LHC) \cite{Aamodt:2010pa,Aamodt:2010pb}.  Among them was a measurement of differential and integrated elliptic flow for unidentified charged hadrons using a variety of methods \cite{Aamodt:2010pa}.  
The differential elliptic flow $v_2$ at a fixed transverse momentum $p_t$ was found to be very similar to the same measurement (using the same method) from lower energy collisions at the Relativistic Heavy Ion Collider (RHIC).  As noted, this was expected from numerous hydrodynamic calculations~\cite{Chojnacki:2007rq,Kestin:2008bh,Niemi:2008ta,Bozek:2010wt} (see also the Appendix).

The values presented for momentum-integrated elliptic flow, however, are larger than most~\cite{Kestin:2008bh,Niemi:2008ta,Chaudhuri:2008je,Luzum:2009sb,Bozek:2010wt} (though not all~\cite{Hirano:2010jg}) hydrodynamic predictions --- at least at first glance.  Compare, for example the ALICE results \cite{Aamodt:2010pa} with the first prediction from a viscous hydrodynamic calculation \cite{Luzum:2009sb}.  Both sets are presented together here in Fig.~\ref{int}.  
%Hirano:2010jg too big

The lower curves represent the published prediction using two different models for hydrodynamic initial conditions (presented as half the momentum anisotropy of the fluid at freeze out, which corresponds closely to the momentum integrated elliptic flow of charged hadrons). 

These predictions were based on an assumed charged hadron multiplicity for a central collision at midrapidity of ${dN_{ch}}/{dY} = 1800$.  This results in a multiplicity per unit pseudorapidity $\ {dN_{ch}}/{d\eta}$ that turned out to be surprisingly close to that seen experimentally (see Appendix).  Thus, if the hydrodynamic model is correct, the integrated elliptic flow should be very close to the experimental value.  

In contrast, both curves are below data for every measurement method at almost every centrality.  This would seem to agree with what is implied in Ref.~\cite{Aamodt:2010pa}, that predictions are too low, unless there is a viscous effect that is reduced between RHIC and LHC.   However, upon closer inspection, it is apparent that the measured quantity is not quite the same as the predicted quantity, and they should not be directly compared.  We will see that a correct comparison results in the conclusion that the integrated elliptic flow at the LHC is just as predicted using the same average viscosity that best fit RHIC data.
\section{Cuts in transverse momentum}
Most theoretical calculations of integrated elliptic flow calculate $v_2$ by averaging over all particles, and Ref.~\cite{Luzum:2009sb} is no exception.  This is an interesting quantity that
contains information about the total momentum anisotropy of the system.
\begin{figure}[t]
\includegraphics[width=\linewidth]{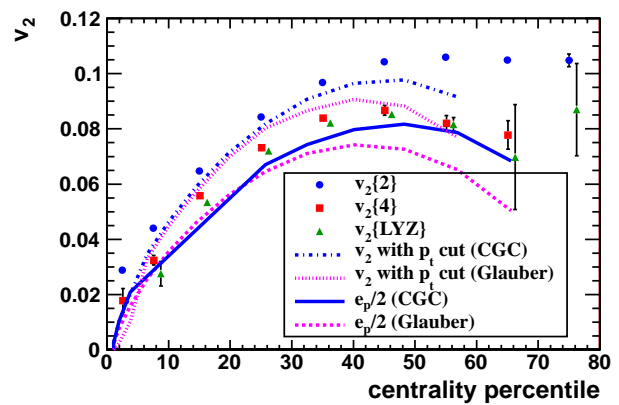}
\caption{(Color online) 
Integrated elliptic flow measurements for charged hadrons using various methods from the ALICE Collaboration (solid symbols \cite{Aamodt:2010pa}) and hydrodynamic predictions (curves).  The two lower curves are predictions using two types of hydrodynamic initial conditions (CGC and Glauber) as presented in Ref.~\cite{Luzum:2009sb}, while the two upper curves represent the integrated $v_2$ for $p_t >200$ MeV from the same hydrodynamic calculation.}
\label{int}
\end{figure}

 ALICE, however, only uses particles with $p_t>200$ MeV.  Excluding these low $p_t$ particles --- which have the smallest elliptic flow --- increases the average elliptic flow. 
The effect of this can be seen by the upper two curves in Fig.~\ref{int}.  Here, using the exact same hydrodynamic evolution as in Ref.~\cite{Luzum:2009sb}, I computed the elliptic flow of charged hadrons with a standard Cooper-Frye prescription and allowing for resonance feed down (see Ref.~\cite{Luzum:2008cw} for details), and then integrated over transverse momentum above 200 MeV.   The cut in momentum increases the integrated value by 14--23\%, depending on centrality. 
%(more than the estimated 12\% in Ref.~\cite{Aamodt:2010pa}).  

This results in predictions that are between the measured values coming from different measurement methods.  Since these methods give significantly different values the next question is precisely to which measurement the predictions should be compared.
\section{Flow analysis methods}% --- flow fluctuations and non flow correlations}
There are two main effects that differentiate between measurement methods --- flow fluctuations and non-flow correlations \cite{Ollitrault:2009ie}.

Non-flow refers to correlations between particles that are independent of the initial geometry of the collision system.  They are expected to provide a positive contribution to measurements of elliptic flow using a two-particle cumulant or event-plane method, but this contribution is significantly reduced in other methods, including 4-particle cumulants, Lee-Yang Zeros, and q-dist \cite{Adler:2002pu}.  They can also be significantly reduced by introducing a rapidity gap \cite{Adler:2003kt}.

Most hydrodynamic calculations, including the one discussed here,  contain no non-flow correlations.

Event-by-event fluctuations of elliptic flow also affect each measurement differently \cite{Alver:2010rt}.
%
%\cite{Alver:2006wh}.  
%
Two-particle cumulant and event-plane methods have a positive contribution while the other methods have a negative contribution.  The sensitivity of the event-plane method to flow fluctuations depends on details of the measurement, but for typical event plane resolutions at RHIC the contribution is only a bit smaller than to two-particle cumulant measurements.  The higher particle cumulants measure a $v_2$ that is similar to the value that would be measured with respect to the reaction plane (as opposed to an event-by-event participant plane which fluctuates around the reaction plane) \cite{Bhalerao:2006tp}.

Our hydrodynamic calculations use smooth initial conditions that represent an average over many events with a fixed reaction plane, and also have no non-flow contribution, so the most appropriate measurements to compare these calculations are the higher order cumulants.

Thus, these predictions are actually a bit too high rather than too low.  However, it turns out that this can be traced to the original RHIC data to which the calculations were fit.  
%The hydrodynamic extrapolation to LHC energy is actually quite well in line with the new measurements.  The only difference is that the best-fit value for shear viscosity that was used is probably slightly too low at both RHIC and LHC (although this difference is much smaller than the uncertainty coming from the eccentricity of the initial conditions.)

\section{RHIC measurements}
The integrated elliptic flow at RHIC was measured by three collaborations, using several methods.  The shear viscosity in our model
%of the hydrodynamic model considered here 
was fixed in Ref.~\cite{Luzum:2008cw} by comparison to an integrated elliptic flow measurement for charged hadrons by PHOBOS (in combination with an estimated higher order cumulant measurement of differential $v_2$ for minimum bias collisions from STAR that was not yet available).  The PHOBOS measurement used an event-plane method with a rapidity gap, and no cut in $p_t$ (although hadrons with $p_t<35$ MeV likely did not make it into the detector) \cite{Alver:2010rt}.  Thus, there was a positive contribution from flow fluctuations, but slightly less than would be expected for a 2-particle cumulant measurement, and most likely very little contribution from non-flow effects.  

These PHOBOS results are similar to higher order cumulant measurements of integrated $v_2$  from STAR, which have a cut in $p_t$ of 150 MeV.  If there were no $p_t$ cut, the STAR results would presumably be lowered such that the PHOBOS result lies between a 2-particle and 4-particle measurement.

Thus, the published prediction for LHC should most likely be expected to fall near the value measured the same way as the one used at RHIC to fix the model parameters --- an event-plane measurement with a rapidity gap and no cut in $p_t$.  In other words, the prediction should lie somewhere between the results of a 2-particle and 4-particle cumulant measurement.  Once the correct cut in $p_t$ is applied, this is exactly what is seen!

The hydrodynamic extrapolation to LHC energy is actually quite well in line with the new measurements.  The only difference is that the best-fit value for shear viscosity that was used is probably slightly too low at both RHIC and LHC (although this difference is much smaller than the uncertainty coming from the eccentricity of the initial conditions, and doesn't affect the conclusions of Ref.~\cite{Luzum:2008cw}).

Note that this means that the implication in Ref.~\cite{Aamodt:2010pa}, that predictions from hydrodynamics will generically be too small unless there is a viscous effect that can be reduced between RHIC and LHC, is not correct.  There is no indication of a significant change in viscosity.

\section{Conclusion}
In conclusion, the new experimental results for elliptic flow at the LHC are just as expected from hydrodynamic models which fix their parameters from RHIC results, indicating another impressive success for hydrodynamics in describing the evolution of a heavy-ion collision.  Precise comparisons require careful attention to the details of each of various measurements, and this will be important in the future when trying to extract more precise quantitative information from experimental data.
\begin{acknowledgments}
I would like to thank R. Snellings for providing the experimental data, as well as P. Romatschke and J.-Y. Ollitrault for helpful comments.  This work was funded by Agence Nationale de la Recherche under grant
ANR-08-BLAN-0093-01.
\end{acknowledgments}
\appendix
\section{Additional results}
\begin{figure}[t]
\includegraphics[width=\linewidth]{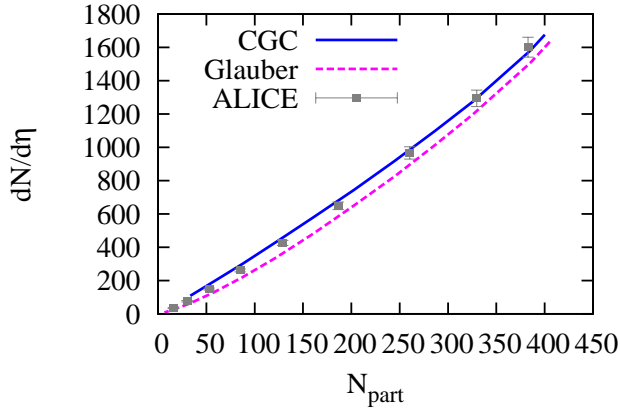}
\caption{(Color online) \label{mult}Charged hadron multiplicity per unit pseudorapidity as a function of the number of participant nucleons $N_{part}$ from viscous hydrodynamics \cite{Luzum:2009sb} along with the experimental result from the ALICE collaboration \cite{Aamodt:2010cz}}
\end{figure}

\begin{figure}[t]
\includegraphics[width=\linewidth]{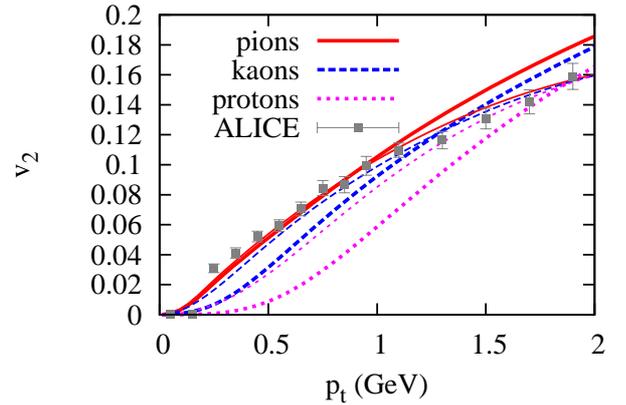}
\caption{(Color online) \label{v2pt}Differential elliptic flow $v_2(p_t)$ for identified particles from viscous hydrodynamic calculations  of LHC collisions at impact parameter $b =$ 7 fm from CGC initial conditions (thick lines \cite{Luzum:2009sb}), along with RHIC calculations at the same impact parameter for comparison (thin lines \cite{Luzum:2008cw}).  Pion elliptic flow (which is essentially equal to charged hadron elliptic flow) at a fixed transverse momentum is unchanged from RHIC to LHC energy, while for heavier particles it is predicted to decrease with collision energy.  Experimental result from the ALICE collaboration is $v_2\{4\}$ for charged hadrons at 20-30\% centrality \cite{Aamodt:2010pa}}
\end{figure}

For the interested reader, additional results from the viscous hydrodynamic calculations are presented---charged hadron multiplicity versus centrality (Fig.~\ref{mult}) and identified particle differential elliptic flow (Fig.~\ref{v2pt}).
\newpage

\end{document}